\documentclass[RNAAS]{aastex62}

\graphicspath{{./}{figures/}}
\newcommand{\tess}{\emph{TESS}}
\begin{document}

\title{Phantom Inflated Planets in Occurrence Rate Based Samples}

\correspondingauthor{L.~C. Mayorga}
\email{laura.mayorga@cfa.harvard.edu}

\author[0000-0002-4321-4581]{L.~C. Mayorga}
\altaffiliation{Harvard Future Faculty Leaders Fellow}
\affiliation{Harvard University}

\author[0000-0002-5113-8558]{Daniel P. Thorngren}
\affiliation{University of California, Santa Cruz}

\keywords{planets and satellites: atmospheres --- planets and satellites: physical evolution --- planets and satellites: gaseous planets --- planets and satellites: detection --- methods: statistical --- surveys}

\section{}
The recently launched \emph{Transiting Exoplanet Survey Satellite} (\tess{}) is expected to produce many new exoplanet discoveries which will be especially amenable to follow-up study.  Assessments of the planet discovery yield of TESS, such as \citet{Sullivan2015} and \citet{Barclay2018}, will be important for planning follow-up work.  Analyzing these predicted planet samples, however, we find that giant planet radii derived from the current bulk transiting planet sample have been used at all potential orbits without accounting for the temperature dependence of radius inflation. The radii of these phantom inflated planets (PIPs) are too large, i.e. beyond the limit of inflation for their equilibrium temperatures. PIP radii should be decreased in accordance with the degree of inflation of the underlying population, however, this may lead to some planets no longer meeting the detectability criteria imposed by yield estimates.

\begin{figure}
\plotone{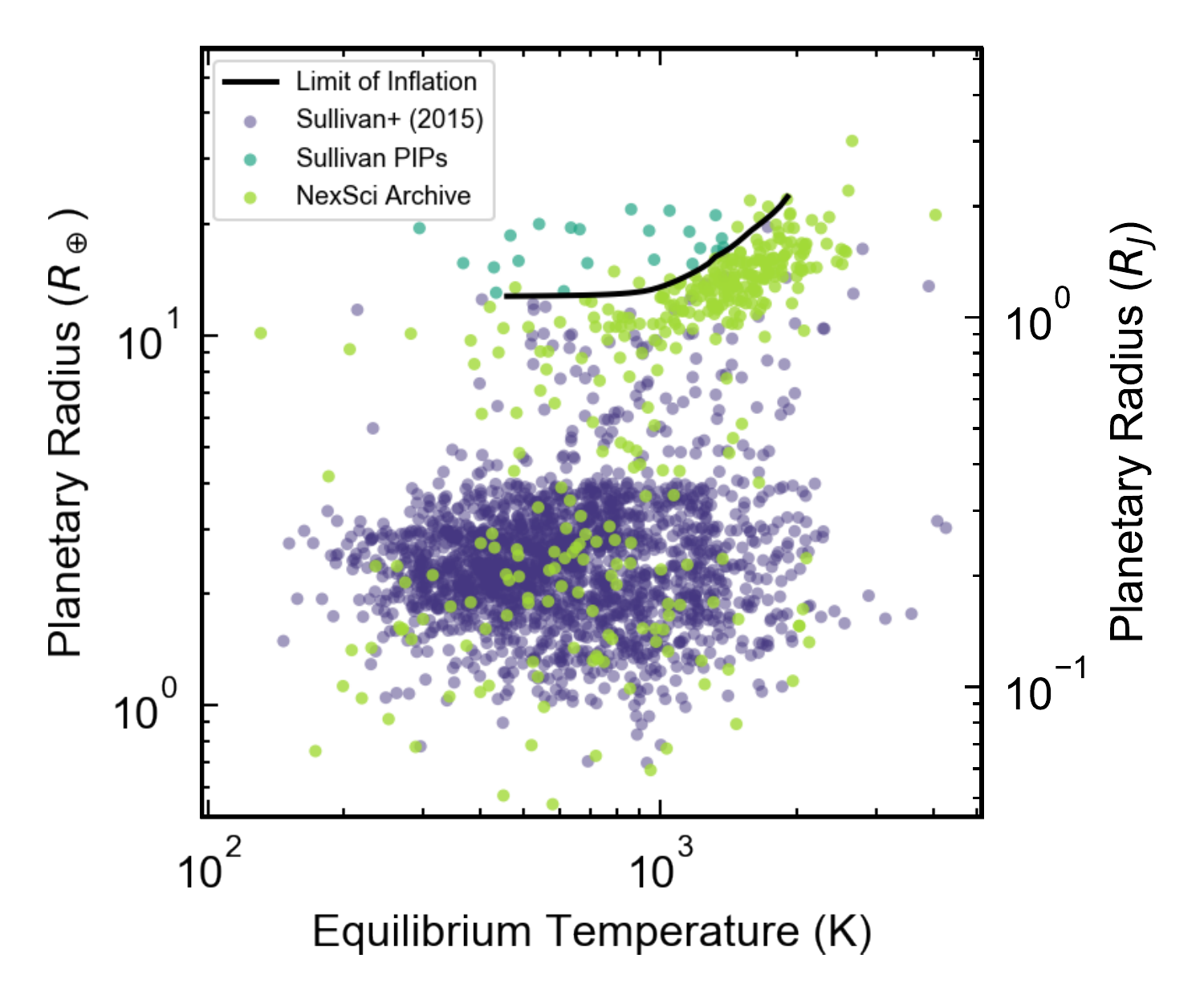}
\caption{The \citet{Sullivan2015} predicted \tess{} sample subdivided by PIP status compared to planets with known equilibrium temperatures and radii from the NASA Exoplanet Archive \citep{Akeson2013} and the limit of inflation  based on \citet{Thorngren2018}.\label{fig:pip}}
\end{figure}

The degree of inflation a planet exhibits is strongly connected to its equilibrium temperature.  Planets below about 1000~K exhibit no significant radius anomaly \citep{Miller2011, DemorySeager2011}, and as such are well-modeled using the same physics as Jupiter and Saturn.  Such planets generally do not exceed $\sim$ 1.2~$R_J$, the radius of a metal-free 1~$M_J$ planet with no anomalous heating \citep{Thorngren2016}.  Beyond 1000~K, radius inflation is correlated with the incident flux \citep{Weiss2013}, or equivalently equilibrium temperature.  For these planets, the radius depends on both the bulk metallicity and the amount of anomalous heat inflating the planet \citep[see e.g.][]{Fortney2010}.  Using the models of anomalous heating from \citet{Thorngren2018}, we can approximate the maximum possible radius of these planets again as a metal-free 1~$M_J$ planet, but now accounting for the inflation effect.  The results are shown as a black line in \autoref{fig:pip} and can be seen to bound nearly all of the observed giants from the NASA Exoplanet Archive \citep{Akeson2013} marked in green.  The few that exceed this limit are either consistent within their error-bars, very young planets, or lower-mass high-flux planets.

Comparing these limits to the predicted yields, we identify a number of PIPs which we mark in teal in \autoref{fig:pip}. PIPs make up just over 1.1\% of the \citet{Sullivan2015} predicted population. For the \citet{Barclay2018} sample PIPs compose about 8\% of the predicted planet population.

It is important to consider the contamination from PIPs for a number of reasons. These PIPs have many characteristics that lead to expectations of easily detectable atmospheres that do not reflect the real population of planets.  Any anticipated follow up science opportunities with this class of planets will not be possible. Similar population predictions for direct imaging studies should likewise take care not to include inflated planets at large separations as such planets will appear larger and, depending on assumed albedo, brighter and more easily detectable than physically possible. Including PIPs in yield simulations for direct imaging missions based on the radial velocity mass distribution would artificially inflate the yield of detected and characterized massive planets. Despite their high false positive rate, giant planets are the testbeds for atmospheric characterization techniques and care should be taken to understand and account for potential contaminating factors in this population.
 
\acknowledgments

\bibliographystyle{aasjournal}
\bibliography{Mendeley}

\begin{thebibliography}{}
\expandafter\ifx\csname natexlab\endcsname\relax\def\natexlab#1{#1}\fi
\providecommand{\url}[1]{\href{#1}{#1}}

\bibitem[{Akeson {et~al.}(2013)Akeson, Chen, Ciardi, Crane, Good, Harbut,
  Jackson, Kane, Laity, Leifer, Lynn, McElroy, Papin, Plavchan, Ram{\'{i}}rez,
  Rey, von Braun, Wittman, Abajian, Ali, Beichman, Beekley, Berriman, Berukoff,
  Bryden, Chan, Groom, Lau, Payne, Regelson, Saucedo, Schmitz, Stauffer, Wyatt,
  \& Zhang}]{Akeson2013}
Akeson, R.~L., Chen, X., Ciardi, D., {et~al.} 2013, Publications of the
  Astronomical Society of the Pacific, 125, 989

\bibitem[{Barclay {et~al.}(2018)Barclay, Pepper, \& Quintana}]{Barclay2018}
Barclay, T., Pepper, J., \& Quintana, E.~V. 2018, 1.
\newblock \url{http://arxiv.org/abs/1804.05050}

\bibitem[{Demory \& Seager(2011)}]{DemorySeager2011}
Demory, B.-O., \& Seager, S. 2011, The Astrophysical Journal Supplement Series,
  197, 12

\bibitem[{Fortney \& Nettelmann(2010)}]{Fortney2010}
Fortney, J.~J., \& Nettelmann, N. 2010, Space Science Reviews, 152, 423

\bibitem[{Miller \& Fortney(2011)}]{Miller2011}
Miller, N., \& Fortney, J.~J. 2011, Astrophysical Journal Letters, 736,
  doi:10.1088/2041-8205/736/2/L29

\bibitem[{Sullivan {et~al.}(2015)Sullivan, Winn, Berta-Thompson, Charbonneau,
  Deming, Dressing, Latham, Levine, McCullough, Morton, Ricker, Vanderspek, \&
  Woods}]{Sullivan2015}
Sullivan, P.~W., Winn, J.~N., Berta-Thompson, Z.~K., {et~al.} 2015, The
  Astrophysical Journal, 809, 77

\bibitem[{Thorngren \& Fortney(2018)}]{Thorngren2018}
Thorngren, D.~P., \& Fortney, J.~J. 2018, The Astronomical Journal, 155, 214

\bibitem[{Thorngren {et~al.}(2016)Thorngren, Fortney, Murray-Clay, \&
  Lopez}]{Thorngren2016}
Thorngren, D.~P., Fortney, J.~J., Murray-Clay, R.~A., \& Lopez, E.~D. 2016, The
  Astrophysical Journal, 831, 64

\bibitem[{Weiss {et~al.}(2013)Weiss, Marcy, Rowe, Howard, Isaacson, Fortney,
  Miller, Demory, Fischer, Adams, Dupree, Howell, Kolbl, Johnson, Horch,
  Everett, Fabrycky, \& Seager}]{Weiss2013}
Weiss, L.~M., Marcy, G.~W., Rowe, J.~F., {et~al.} 2013, The Astrophysical
  Journal, 768, 14

\end{thebibliography}

\end{document}